\documentclass[preprint,showpacs,preprintnumbers,amsmath,amssymb]{revtex4}
\usepackage{enumerate}
\usepackage[dvips]{graphicx}
\usepackage{bm}
\begin{document}
\preprint{Preprint}
\title{Critical dynamics of phase transition driven by \\
 dichotomous Markov noise}
\author{Katsuya Ouchi}
 \email{ouchi@kobe-du.ac.jp}
 \affiliation{Kobe Design University, 8-1-1 Gakuennishi-machi,
 Kobe 651-2196, Japan}
\author{Takehiko Horita}
\altaffiliation[Present address: ]{Department of Mathematical Sciences,
Osaka Prefecture University, 1-1 Gakuencho, Sakai, Osaka 599-8531, Japan}
 \email{horita@ms.osakafu-u.ac.jp}
 \affiliation{Department of Mathematical Informatics,
 The University of Tokyo, 7-3-1 Hongo, Bunkyo-ku, Tokyo 113-8656, Japan}
\author{Hirokazu Fujisaka}
 \email{fujisaka@i.kyoto-u.ac.jp}
 \affiliation{Department of Applied Analysis and Complex Dynamical Systems,
 Graduate School of Informatics,
 Kyoto University, Kyoto 606-8501, Japan}
\date{\today}
\begin{abstract}
 An Ising spin system under the critical temperature
 driven by a dichotomous Markov noise (magnetic field) with a finite
 correlation time is studied both numerically and theoretically.
 The order parameter exhibits a transition between two kinds of
 qualitatively different dynamics, \textit{symmetry-restoring}
 and \textit{symmetry-breaking motions},
 as the noise intensity is changed.
\par
 There exist regions called channels where the order parameter stays
 for a long time slightly above its critical noise intensity.
 Developing a phenomenological analysis of the dynamics, we investigate the
 distribution of the passage time through the channels and the power spectrum
 of the order parameter evolution.
 The results based on the phenomenological analysis turn out to be in quite
 good agreement with those of the numerical simulation.
\end{abstract}
\pacs{64.60.-i,05.40.-a,02.50.-r,05.70.Jk}
\maketitle
\section{Introduction}
 Over the last decade, the dynamics of ferromagnetic systems
 below their critical temperatures in a periodically oscillating
 magnetic field have been studied both
 theoretically~\cite{tome,lo,acharyya,sides,rikvold,korn1,fujisaka1,yasui,tutu}
 and experimentally~\cite{jiang}.
 The systems exhibit two qualitatively different
 behaviors referred to as \textit{symmetry-restoring oscillation} (SRO)
 and \textit{symmetry-breaking oscillation} (SBO), depending on
 the frequency \(\Omega\) and the amplitude \(h\) of the applied
 magnetic field.
 It has been established that there exists a sharp transition line
 between SRO and SBO on the (\(\Omega\), \(h\)) plane,
 which is called the \textit{dynamical phase transition} (DPT).
 The DPT was first observed numerically in the deterministic mean-field
 system for a ferromagnet in a periodically oscillating field~\cite{tome},
 and has subsequently been studied in numerous Monte Carlo simulations
 of the kinetic Ising system below critical
 temperature~\cite{lo,acharyya,sides,rikvold,korn1}.
 It has also been observed experimentally in an ultra-thin Co film on
 Cu(100)~\cite{jiang}.
\par
 Recently, we investigated the DPT by introducing the
 model equation
 $\dot{s}(t) = (T_c-T) s - s^3 + h\cos \Omega t$~\cite{fujisaka1}.
 This equation is a simplified  model for the Ising spin system at the
 temperature $T$ below its critical value $T_c$ in an external periodic
 magnetic field.  By appropriately scaling the magnetization $s$, time $t$,
 and the applied field, this equation is written as
\begin{equation}
\dot{s}(t) = s - s^3 + h \cos \Omega t.
\label{eq:tdgl-scale-external-field}
\end{equation}
 The SBO and SRO are observed in
 Eq.~(\ref{eq:tdgl-scale-external-field})
 and the transition line between them on the (\(\Omega\), \(h\)) plane
 is determined analytically~\cite{fujisaka1,tutu}.
\par
 It is quite interesting to ask whether DPT is observed
 under another kind of applied field,
 especially random field with bounded amplitude.
 The fundamental aim of the present paper is to study the dynamics of \(s(t)\)
 with a dichotomous Markov noise (DMN) \(F(t)\) instead of periodically
 oscillating external field \(h \cos \Omega t\) (see, e.g.,~\cite{kampen}).
\par
 The equation of motion
\begin{equation}
\dot{s} = f(s) + F(t)
\label{eq:general-external-field}
\end{equation}
with a nonlinear function \(f(s)\) and the DMN \(F(t)\) has been
extensively studied by many
authors~\cite{kitahara,bena,heureux,hanggi}.
 It is well known that the master equation for the system can be
 derived, and then transition phenomena of stationary probability
 densities concerning the intensity of \(F(t)\), for example, are
 studied, which are referred to as the noise-induced phase
 transition~\cite{kitahara,horsthemke}.
 The asymptotic drift velocity \(\langle \dot{s} \rangle\) in the case
 of \(f(s)\) being periodic functions are also discussed as a specific
 dynamic property~\cite{bena}.
 Furthermore, the mean first-passage time (MFPT) and transition rates
 are investigated as another important dynamic property when \(f(s)\)
 is the force associated with the bistable potential given by
 Eq.~(\ref{eq:general-external-field})~\cite{heureux,hanggi}.
 For a review of works on DMN system, see Bena~\cite{bena2}.
\par
 The fundamental aim of the present paper is to propose a
 phenomenological approach to the critical dynamics near the 
 transition point between the symmetry-restoring motion (SRM) 
 and the symmetry-breaking motion (SBM) observed in
 Eq.~(\ref{eq:general-external-field}) with \(f(s)=s-s^3\)
 (Sec.~\ref{sec:model-equation}) and to investigate the
 distribution of passage times through channels, switching times
 between two different motions, and the power spectrum of the order
 parameter evolution.
 The present paper is constructed as follows.
 In Sec.~\ref{sec:model-equation}, we discuss the dynamics of
 symmetry-breaking motion and symmetry-restoring motion
 of the model equation (\ref{eq:general-external-field}).
 In Sec.~\ref{sec:average-passage-time}, the jumping process of the
 magnetization through channels, which are defined in the text, is
 investigated and the MFPT is obtained.
 In Sec.~\ref{sec:probabilistic-analysis}, a phenomenological approach
 simplifying
 the dynamics of passing through a channel in the SRM phase is introduced
 and three statistical characteristics are analytically developed.
 The results are compared with numerical simulations.
 Concluding remarks are given in Sec.~\ref{sec:concluding}.
%
%
%
%
\section{Model equation and symmetry-breaking transition}
\label{sec:model-equation}
\subsection{Model equation and noise-induced phase transition}
\label{sec:model-equation-and-basic-property}
\par
 We consider the equation of motion driven by the external field \(F(t)\),
\begin{equation}
\frac{d s(t)}{dt} = f(s) + F(t), \quad (f(s)=s-s^3)
\label{eq:tdgl-external-field}
\end{equation}
where \(F(t)\) is a symmetric DMN with taking the values \(\pm H_0\).
 Here the probability \(p(\tau)\) that \(F(t)\) continues to take
 the identical value \(+H_0\) or \(-H_0\) longer than time \(\tau\)
 is given by
\begin{equation}
p(\tau) = e^{-\tau / \tau_f}.
\label{eq:field-probability}
\end{equation}
 This implies that the correlation time of \(F(t)\) is equal to \(\tau_f / 2\).
 Throughout this paper, numerical integrations of
 Eq.~(\ref{eq:tdgl-external-field}) are carried out by using the Euler
 difference scheme with the time increment \(\Delta t = 1/100\).
\par
\begin{figure}
\includegraphics[scale=0.5]{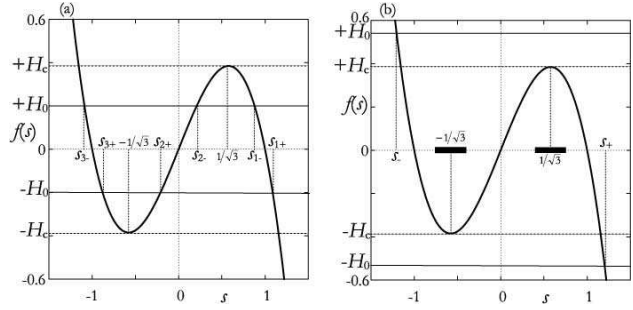}
\caption{\label{fig:f_x} The function \(f(s)=s-s^3\) is shown by solid lines.
 (a) Real roots \(s_{j \pm}\), (\(j=1\), 2, 3)
 for \(H_0 < H_c\) and (b) real roots \(s_{\pm}\) for \(H_0 > H_c\)
 of the algebraic equation \(s - s^3 \pm H_0 = 0\).
 The definition of \(H_c\) is graphically represented.
 Two bold lines drawn in (b) indicate the channel regions
 defined in Sec.~\ref{sec:average-passage-time}.
}
\end{figure}

 Without DMN, \(s(t)\) eventually approaches
 either of the stationary fixed points \(\pm 1\),
 one of which is achieved according to the initial condition \(s(0)\)
 as shown in Fig.~\ref{fig:f_x}.
 In the presence of DMN, if \(H_0 < H_c\), \(H_c\) being defined by
\begin{equation}
H_c \equiv 2 (1/3)^{3/2} = 0.3849 \cdots,
\label{eq:define_Hc}
\end{equation}
 then \(f(s) + H_0=0 (f(s) - H_0=0)\) has three
 real roots \(s_{j+}(s_{j-})\), (\(j=1\), 2, and 3).
 Each value of \(s_{j\pm}\) is graphically shown in Fig.~\ref{fig:f_x}(a).
 On the other hand, if \(H_0 > H_c\),
 then \(f(s) + H_0 = 0\) (\(f(s) - H_0 = 0\)) has only one real root
 \(s_{+}\) (\(s_{-}\)) given by
\begin{equation}
s_{\pm} =
 \left[ \frac{1}{2} \left( \pm H_0 + \sqrt{H_0^2 - H_c^2}\right) \right]^{1/3}
+\left[ \frac{1}{2} \left( \pm H_0 - \sqrt{H_0^2 - H_c^2}\right) \right]^{1/3},
\label{eq:define-s_pm}
\end{equation}
 which are indicated in Fig.~\ref{fig:f_x}(b).
\par
\begin{figure}
\includegraphics[scale=0.5]{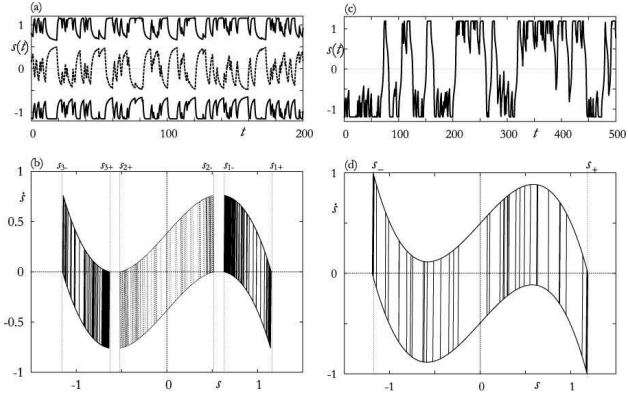}
\caption{\label{fig:orb} Figures (a) and (b) show the motions obtained
 by numerically integrating Eq.~(\ref{eq:tdgl-external-field})
 for \(H_0 = 0.38 (< H_c)\) and \(\tau_f = 5\),
 where two SBM's (solid line) and an unstable SRM (dashed line) are drawn.
 The unstable SRM is evaluated by replacing \(t \to -t\).
 On the other hand, Figs.~(c) and (d) show the motions
 for \(H_0 = 0.5 (> H_c)\) and \(\tau_f = 10\).
}
\end{figure}
 Next let us consider the dynamics described by
 Eq.~(\ref{eq:tdgl-external-field}) for \(H_0 < H_c\) and for \(H_0 > H_c\),
 and discuss similarity and difference between
 the dynamics in the periodically oscillating field case and those
 in the present DMN case.
 A part of our results belongs to the context of the noise-induced
 phase transition and MFPT in
 Refs.~\cite{kitahara,horsthemke,bena,heureux,hanggi}.
 In the case of \(H_0 < H_c\), three motions numerically integrated
 are shown in Figs.~\ref{fig:orb}(a) and (b).
 Two motions confined in the ranges
 \(s_{1-} < s(t) < s_{1+}\) and \(s_{3-} < s(t) < s_{3+}\) are both stable.
 The long time average \(\langle s(t) \rangle\) of each motion does not vanish,
 and the motion is called SBM in relation to DPT in the oscillating
 external field case.
 On the other hand, the motion \(s_u(t)\) confined in the range
 \(s_{2+} < s_u(t) < s_{2-}\) is unstable.
 The long time average of \(s_u(t)\) vanishes, and in this sense
 the motion is called SRM.
 It should be noted that this unstable SRM
 is located between two stable SBM, which has a similar characteristic
 to SBO of DPT~\cite{fujisaka1}.
\par
 The motion of \(s(t)\) for \(H_0 > H_c\) is shown in
 Figs.~\ref{fig:orb}(c) and (d).
 One observes that there exists a stable SRM confined in the range
 \(s_{-} < s(t) < s_{+}\).
 For SRM, the time average of \(s(t)\) vanishes,
 i.e., \(\langle s(t) \rangle = 0\).
 The comparison between Figs.~\ref{fig:orb}(b) and (d)
 suggests that the SRM for \(H_0>H_c\) is generated via the ``attractor
 merging crisis''~\cite{ott} of the two SBM's and one unstable SRM, i.e.,
 the two SBM's and one unstable SRM disappear and then
 one stable SRM takes place at \(H_0 = H_c\).
 This situation is similar to that in the DPT case.
 However, in contrast to the DPT case, as will be
 shown in Sec.~\ref{sec:stationary-distribution}, the transition line
 on the (\(\tau_f^{-1}\), \(H_0\)) plane is
 independent of the correlation time \(\tau_f\) of \(F(t)\)
 and the average \(\langle s(t) \rangle\)
 depends discontinuously on \(H_0\).
\subsection{Stationary distribution functions and phase diagram}
\label{sec:stationary-distribution}
\par
 In this subsection, we discuss the stationary distribution functions
 for SBM and SRM.
 To this aim, we first consider a slightly general nonlinear Langevin
 equation of motion driven by DMN,
\begin{equation}
\dot{x}(t) = f(x) + g(x)F(t),
\label{eq:general-stochastic}
\end{equation}
where \(f(x)\) and \(g(x)\) are generally nonlinear functions of \(x\)
 and \(F(t)\) is DMN~\cite{markov}.
 The temporal evolution of
 the distribution function \(P(x, F, t)\) that \(x(t)\) and \(F(t)\)
 respectively take the values \(x\) and \(F (= \pm H_0)\)
 is determined by \cite{kitahara,horsthemke}
\begin{eqnarray}
\frac{\partial}{\partial t}P(x,t) &=&
 - \frac{\partial}{\partial x} \left[ f(x) P(x,t) + H_0 g(x) q(x,t) \right],
 \nonumber \\
\frac{\partial}{\partial t}q(x,t) &=& - \frac{2}{\tau_f} q(x, t)
 -  \frac{\partial}{\partial x} \left[ f(x)q(x,t) + H_0 g(x)P(x,t) \right],
\label{eq:general-probability-evolution}
\end{eqnarray}
where we put \(P(x,t) \equiv P(x, + H_0,t) + P(x, -H_0,t)\) and
 \(q(x,t) \equiv P(x, +H_0,t) - P(x, -H_0,t)\).
 The stationary distribution \(P^{st}(x) \equiv P(x, \infty)\)
 is solved to yield
\begin{equation}
P^{st}(x) = N \frac{g(x)}{H_0^2 g(x)^2 - f(x)^2}
 \exp \left\{ - \frac{1}{\tau_f} \int^{x} dx'
\left[ \frac{1}{f(x')-H_0 g(x')} + \frac{1}{f(x')+H_0 g(x')} \right] \right\},
\label{eq:general-stationary-distribution}
\end{equation}
provided that each of the equations
\begin{equation}
\dot{x} = f(x) + H_0 g(x), \quad \dot{x} = f(x) - H_0 g(x)
\end{equation}
has at least one stable fixed point, where \(N\) is the
 normalization constant.
\par
 By substituting
 \(f(x)=x-x^3\) and \(g(x)=1\), (Eq.~(\ref{eq:tdgl-external-field})), into
 Eq.~(\ref{eq:general-stationary-distribution}),
 the stationary distribution function \(P_{SBM}^{st}(s)\)
 for SBM (\(H_0 < H_c\)) for \(s_{3-} < s < s_{3+}\) or
 \(s_{1-} < s < s_{1+}\) is written as
\begin{eqnarray}
P_{SBM}^{st}(s) &\propto& |s^2 - s_{1+}^2|^{- \beta_{1+}}
 |s^2 - s_{1-}^2|^{- \beta_{1-}} |s^2 - s_{2+}^2|^{- \beta_{2+}},
\label{eq:distribution-sbm} \\
 \beta_{j \pm} &=& 1 - \tau_f^{-1} |(s_{j\pm}-s_{k\pm})(s_{j\pm}-s_{l\pm})|,
\end{eqnarray}
where \((j,k,l)=(1,2,3)\), (2,3,1), and (3,1,2).
\begin{figure}
\includegraphics[scale=0.5]{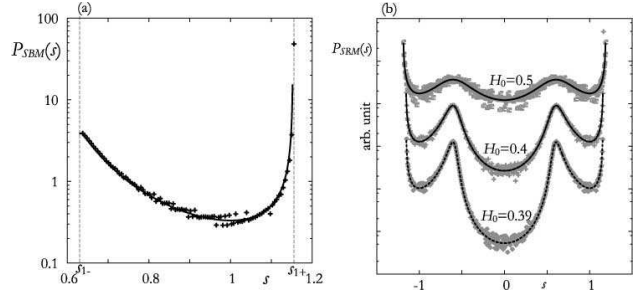}
\caption{\label{fig:distrib} Stationary distribution functions
 obtained both analytically (solid line) and numerically (symbols)
 for (a) \(H_0 = 0.3\) and \(\tau_f = 5\),
 and for (b) \(H_0=0.39\), 0.4, and 0.5 by keeping \(\tau_f = 10\).
 (a) and (b) correspond respectively to SBM and SRM.
 The distribution only in the range \(s_{1-} < s(t) < s_{1+}\) is drawn
 in (a).
 The analytical result is in good agreement with
 that of the numerical simulation.
}
\end{figure}
 On the other hand, the stationary distribution function
 \(P_{SRM}^{st}(s)\) for the SRM (\(H_0 > H_c\)) for  \(s_{-} < s < s_{+}\)
 is obtained as
\begin{eqnarray}
P_{SRM}^{st}(s) &\propto& | s^2 - s_{+}^2 |^{\frac{\tau_f^{-1}}{3 s_{+}^2 -1}
 - 1} \left[ (s^2 + s_{+}^2 - 1)^2 - s_{+}^2 s^2 \right]^{
- \frac{\tau_f^{-1}}{3 s_{+}^2 -1} - 1} \nonumber \\
&\times& \exp \Bigg\{\frac{\tau_f^{-1} s_{+}}{(s_{+}^2-1/3)\sqrt{3s_{+}^2-4}}
\Bigg[ \arctan \left( \frac{2 s - s_{+}}{\sqrt{3 s_{+}^2 - 4}} \right) \nonumber \\
& & - \arctan \left( \frac{2 s + s_{+}}{\sqrt{3 s_{+}^2 - 4}} \right) \Bigg]
 \Bigg\}.
\label{eq:distribution-srm}
\end{eqnarray}
 The analytic solutions (\ref{eq:distribution-sbm}) and
 (\ref{eq:distribution-srm}) are numerically confirmed in
 Fig.~\ref{fig:distrib}.
\par
\begin{figure}
\includegraphics[scale=0.5]{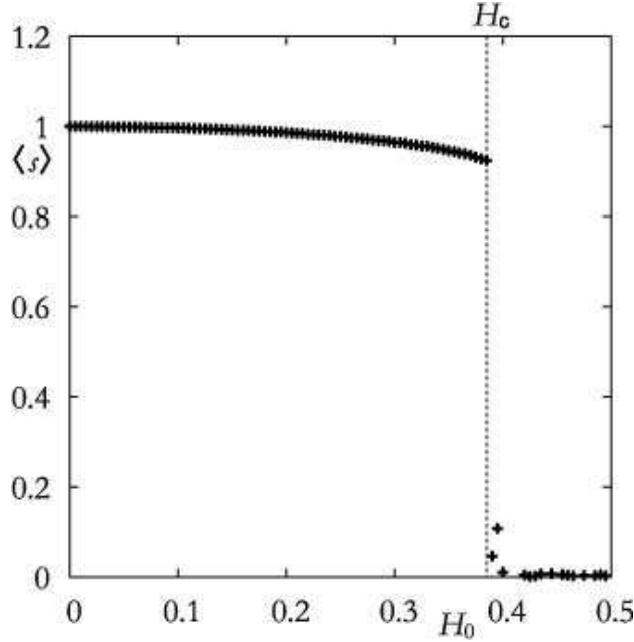}
\caption{\label{fig:disconnect} \(\langle s \rangle\) vs \(H_0\)
 for \(\tau_f = 5\).
 One finds that \(\langle s \rangle\) is discontinuous at \(H_0 = H_c\).
}
\end{figure}

 As \(H_0\) is increased,
 the form of the stationary distribution function
 changes drastically from the forms in Eq.~(\ref{eq:distribution-sbm}) to
 Eq.~(\ref{eq:distribution-srm}) at \(H_0 = H_c\).
 This phenomenon which is induced by the disappearance of two pairs
 of stable and unstable fixed points~\cite{bena} is an example of the
 noise-induced phase transitions~\cite{horsthemke}.
\begin{figure}
\includegraphics[scale=0.5]{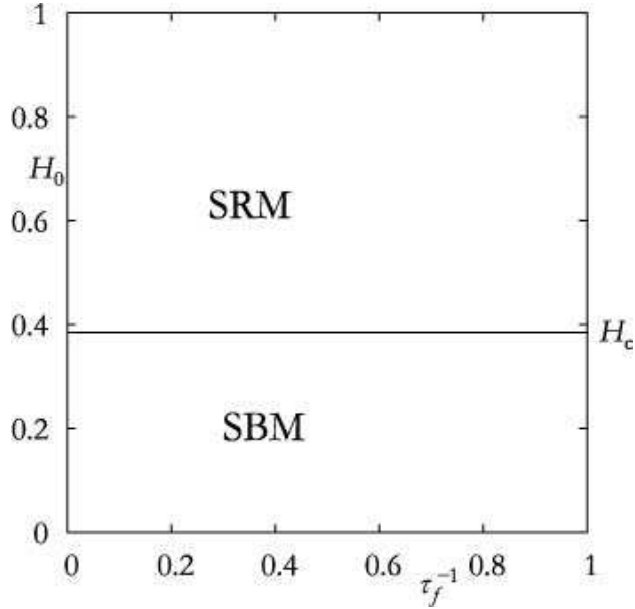}
\caption{\label{fig:phase_diagram} Phase diagram on the
 (\(\tau_f^{-1}\), \(H_0\)) plane.
 Transition line is independent of \(\tau_f^{-1}\)
 and is given by \(H_0 = H_c\).
}
\end{figure}
 It turns out that the transition line between SRM and SBM on the
 (\(\tau_f^{-1}\), \(H_0\)) plane is given
 by \(H_0 = H_c\).
 The phase diagram is given in Fig.~\ref{fig:phase_diagram}.
 Furthermore, the long time average of \(s(t)\), \(\langle s(t) \rangle\),
 depends discontinuously on \(H_0\) at \(H_0 = H_c\) as shown
 in Fig.~\ref{fig:disconnect}.
 These behaviors are quite different from those of
 the DPT case driven by periodically oscillating
 field, \(F(t)=h \cos (\Omega t)\)~\cite{acharyya,fujisaka1}.
 The transition point \(h_c\) for a fixed \(\Omega\) between SRO and SBO
 depends on the frequency
 \(\Omega\), and \(\langle s(t) \rangle\) is a continuous function of \(h\).
\section{MFPT through the channels}
\label{sec:average-passage-time}
 We hereafter discuss the dynamics for \(H_0\) slightly
 above \(H_c\).
 Let us first consider the behavior obeying the equations
\begin{equation}
\dot{s} = s - s^3 + \epsilon H_0, \quad (\epsilon = + \, \mathrm{or} \, -)
\label{eq:tdgl-fixed-field}
\end{equation}
 for \(H_0 > H_c\), i.e., \(F(t)\) is fixed to be
 either \(+H_0\) or \(-H_0\).
 Equation~(\ref{eq:tdgl-fixed-field}) for \(H_0 > H_c\) is
 integrated to yield
\begin{eqnarray}
t &=& - \frac{1}{2(3 s_{\epsilon}^2-1)} \ln
 \frac{(s-s_{\epsilon})^2}{s^2+s_{\epsilon}
 s+s_{\epsilon}^2-1} \frac{s_0^2+s_{\epsilon}
 s_0+s_{\epsilon}^2-1}{(s_0-s_{\epsilon})^2} \nonumber\\
& & + \frac{6 s_{\epsilon}}{2(3s_{\epsilon}^2-1)
 \sqrt{3 s_{\epsilon}^2-4}} \left[
 \arctan \left( \frac{2s+ s_{\epsilon}}{\sqrt{3 s_{\epsilon}^2-4}} \right)
 - \arctan \left( \frac{2s_0+ s_{\epsilon}}{\sqrt{3 s_{\epsilon}^2-4}}
 \right) \right],
\label{eq:tdgl-srm-motion}
\end{eqnarray}
where \(s_0 = s(0)\) and \(s_{\epsilon}\) has been defined
 in Eq.~(\ref{eq:define-s_pm}).
\begin{figure}
\includegraphics[scale=0.5]{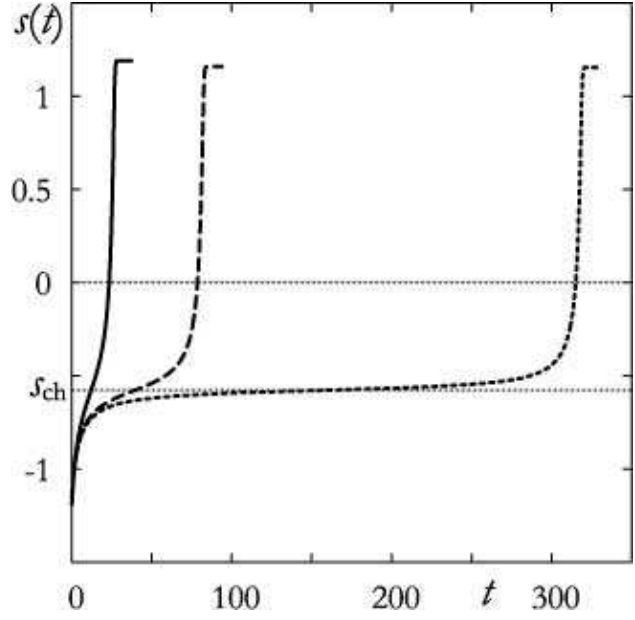}
\caption{\label{fig:anal_sro_orbit} Three orbits of the equation of motion
 (\ref{eq:tdgl-fixed-field}) with \(\epsilon = +\) for \(H_0 > H_c\).
 The values of \(H_0\) are set to be \(H_0=0.386\) (dotted line),
 0.4 (dashed line), and 0.5 (solid line),
 where all the initial conditions are chosen as \(s_0 = s_{-}\).
}
\end{figure}
 Figure~\ref{fig:anal_sro_orbit} displays three orbits given by
 Eq.~(\ref{eq:tdgl-srm-motion}) with \(s_0 = s_{-}\) and \(\epsilon = +\),
 which shows that \(s(t)\) approaches \(s_{+}\) in the limit
 \(t \to \infty\).
 One observes that \(s(t)\) stays for a long time
 in the vicinity of \(s = -1/\sqrt{3}\) for \(H_0\) slightly above
 \(H_c\).
 The small region including the position \(s=-1/\sqrt{3}\)
 is called the `channel'.
 From the symmetry of the system,
 there also exists the channel near \(s=1/\sqrt{3}\) for \(F(t)=-H_0\),
 as shown in Fig.~\ref{fig:f_x}(b).
 Let us express the positions \(s_{ch}\) of the channels as
\begin{equation}
s_{ch} =
\left\{
\begin{array}{cc}
 -1/\sqrt{3}, & \mathrm{if} \, F(t)=+H_0 \\
 +1/\sqrt{3}, & \mathrm{if} \, F(t)=-H_0
\end{array}
\right. .
\end{equation}
\par
 The characteristic time \(\tau_{ch}\) is then defined as the time span that
 the state point \(s(t)\) passes through one of the channels
 for a constant \(F(t)\),
 either \(+ H_0\) or \(- H_0\).
 \(\tau_{ch}\) can be estimated by integrating
 Eq.~(\ref{eq:tdgl-fixed-field}) around \(s \simeq s_{ch}\) as follows.
 First, consider the case \(F(t)= - H_0\).
 By setting \(u(t) = s(t) - s_{ch}\) and assuming \(|u| \ll s_{ch}\),
 Eq.~(\ref{eq:tdgl-fixed-field}) is approximated as
\begin{equation}
\dot{u} = - 3 s_{ch} u^2 - (H_0 - H_c).
\end{equation}
 This can be integrated to give
\begin{equation}
u(t) = - \sqrt{ \frac{H_0 - H_c}{3 s_{ch}}} \tan
 \left[ \sqrt{3 s_{ch} (H_0 - H_c)} t \right]
\label{eq:channel-motion}
\end{equation}
with the initial condition \(u(0)=0\).
 \(\tau_{ch}\) is estimated by the condition \(u(\tau_{ch}) = \infty\) and
 thus
\begin{equation}
\tau_{ch} = \frac{C}{(H_0 - H_c)^{1/2}},
 \quad C = \frac{\pi}{2 \sqrt{3 s_{ch}}}.
\label{eq:tau_channel}
\end{equation}
\par
\begin{figure}
\includegraphics[scale=0.5]{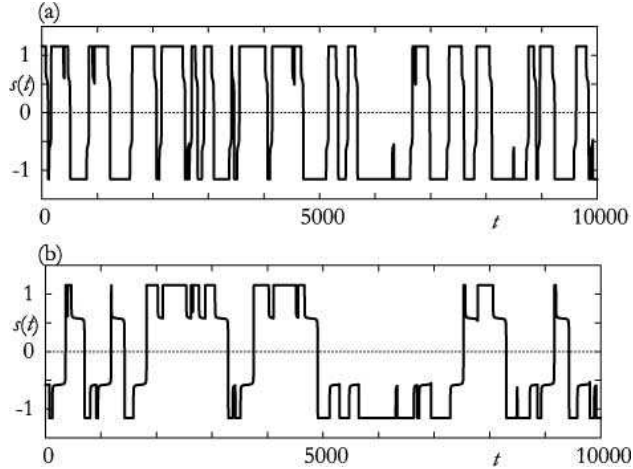}
\caption{\label{fig:time-series} Time series of \(s(t)\) for
 (a) \(H_0=0.388\) and (b) 0.385 by keeping \(\tau_f= 200\).
}
\end{figure}

 Let us next consider the process that the state point \(s(t)\)
 passes through the channels under DMN.
 Figure~\ref{fig:time-series} shows temporal evolutions
 of \(s(t)\) numerically obtained for \(H_0 = 0.388\) and 0.385.
 One finds that the time of passing through
 channels increases as \(H_0\) approaches \(H_c\).
 The MFPT \(\bar{\tau}\) through channels was calculated
 in Refs.~\cite{heureux,hanggi} by analyzing the master equation.
 In the present section, we will derive MFPT in terms of the time
 scales \(\tau_f\) and \(\tau_{ch}\) from a phenomenological
 viewpoint without use of the analysis made in Refs.~\cite{heureux,hanggi}.
\par
 The condition for passing through a channel
 is that \(F(t)\) continues to take the identical value either \(+H_0\)
 or \(- H_0\) for time longer than \(\tau_{ch}\).
 For \(H_0\) satisfying \(\tau_f > \tau_{ch}\),
 we obtain \(p(\tau_{ch})= e^{-\tau_{ch}/\tau_{f}} \simeq 1\), which
 implies that \(F(t)\) almost always satisfies the condition
 for passing through the channel.
 Therefore, \(\bar{\tau}\) in the case of \(\tau_f > \tau_{ch}\)
 is nearly equivalent to \(\tau_{ch}\), i.e.,
\begin{equation}
 \bar{\tau} \simeq \frac{C}{(H_0 - H_c)^{1/2}}.
\label{eq:average-time-far-point}
\end{equation}
\par
 In the case of \(\tau_f \ll \tau_{ch}\), on the other hand,
 Eq.~(\ref{eq:field-probability}) gives
 \(p(\tau_{ch}) \ll 1\).
 This fact implies that the probability that \(F(t)\)
 continues to take the identical value for time longer than \(\tau_{ch}\) is
 quite small and hence that \(\bar{\tau}\) is much longer
 than \(\tau_{ch}\) because it needs a long time to satisfy the condition
 for the state point to pass through the channel.
 \(\bar{\tau}\) in the case of \(\tau_f \ll \tau_{ch}\) is explicitly
 determined as follows.
 For a long \(\bar{\tau}\), let us divide \(\bar{\tau}\) into
 subintervals each of which has the time span \(\tau_f\).
 The divided individual time series are approximately
 independent of each other.
 Therefore, \(\tau_f /\bar{\tau}\) is the probability that the state point
 passes through a channel once because \(\bar{\tau}\) is MFPT
 through the channel.
 On the other hand, \(p(\tau_{ch})\) is identical to the
 probability for \(s(t)\)
 to pass through the channel once by definition of the probability.
 Therefore we get the relation
 \(p(\tau_{ch}) \simeq \tau_f / \bar{\tau}\), which leads to
\begin{equation}
\bar{\tau}^{-1} \simeq  \tau_f^{-1} e^{- \tau_{ch}/\tau_f} =
 \tau_f^{-1} \exp \left[ - \frac{C}{\tau_f ( H_0 - H_c )^{1/2}} \right]
\label{eq:estimate-tau-vs-h0}
\end{equation}
with the constant \(C\) defined in Eq.~(\ref{eq:tau_channel}).
This expression agrees with the result obtained
 in Refs.~\cite{hanggi, heureux}.
\begin{figure}
\includegraphics[scale=0.5]{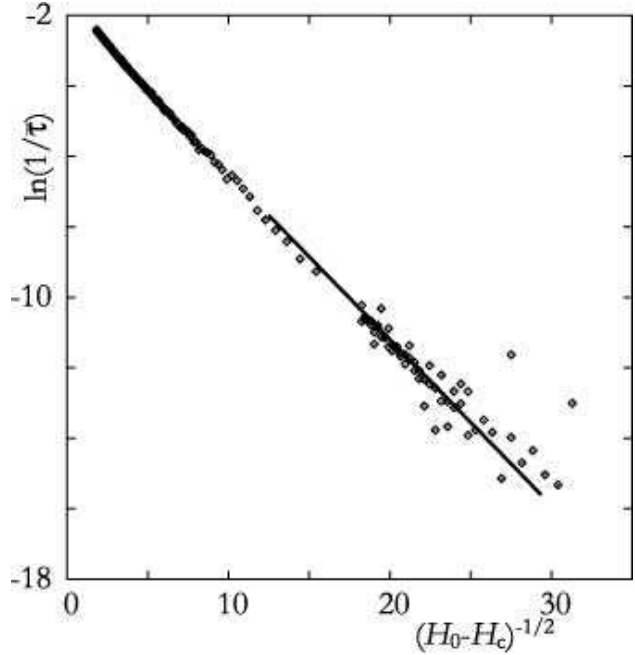}
\caption{\label{fig:tau_H0} The dependence of \(\bar{\tau}\) on \(H_0 - H_c\)
 determined numerically for \(\tau_f = 5\).
 \(\bar{\tau}\) for a given \(H_0\) is obtained by integrating
 Eq.~(\ref{eq:tdgl-external-field}) until a time \(T\)
 and simultaneously by counting the number \(N\) of times passing through
 the position \(s=0\).
 Then \(\bar{\tau}\) is evaluated as
 \(\bar{\tau} = T/N\).
 One finds that \(\ln \bar{\tau}^{-1}\) linearly depends on
 \((H_0 - H_c)^{-1/2}\) in the region of \(0 < H_0 - H_c \ll 1\),
 where the solid line is expressed as \(\ln y = A x + B\) with
 fitting coefficients \(A\) and \(B\).
}
\end{figure}
 Equation~(\ref{eq:estimate-tau-vs-h0}) reveals that MFPT
 through the channel depends on \(H_0-H_c\)
 in a stretched exponential form for \(\tau_f \ll \tau_{ch}\), and is
 quite different from the asymptotic form (\ref{eq:average-time-far-point}).
 The above dependence of \(\bar{\tau}\) on \(H_0 - H_c\) is confirmed
 in Fig.~\ref{fig:tau_H0}.
\section{Phenomenological Analysis}
\label{sec:probabilistic-analysis}
\begin{figure}
\includegraphics[scale=0.5]{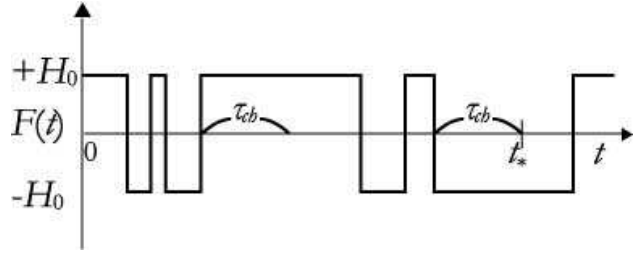}
\caption{\label{fig:jump_schematic} Time series of \(F(t)\)
 schematically indicating the time that \(s(t)\) jumps from \(s_+\)
 to \(s_-\).
 \(s(t)\) jumps at the time \(t_{*}\) in this case,
 where \(F(t)\) takes the value \(-H_0\) longer than \(\tau_{ch}\)
 for the first time in the time series of \(F(t)\).
}
\end{figure}
 In order to discuss statistical characteristics of the dynamics
 passing through the channels for \(\tau_f \ll \tau_{ch}\),
 we here develop a phenomenological approach.
 The behaviors of \(s(t)\) for which we attempt to model are first summarized.
 The initial condition of \(s(t)\) is set to be in the vicinity of \(s_+\).
 If a time interval of \(F(t)\) satisfying the condition \(F(t) = - H_0\)
 becomes longer than \(\tau_{ch}\) for the first time,
 then \(s(t)\) passes through \(s_{ch}\)
 and approaches \(s_-\) in the time interval.
 See Fig.~\ref{fig:jump_schematic}.
The event in which \(s(t)\) jumps from \(s_+\) to \(s_-\) occurs only
 in this case.
 It should be noted that the jumps from \(s(t)>0\) (\(s(t)<0\)) to
 \(s(t)<0\) (\(s(t)>0\)) are approximately independent of subsequent jumps.
\par
 Let us discretize the time \(t\) in the form \(t = k \Delta t\),
 (\(k=1\), 2, 3, \(\cdots\))
 as a simple approach to develop the phenomenological analysis
 according to the process noted above, where \(\Delta t\) is a certain
 small time step.
 Then, \(\tau_{ch}\) is discretized as \(\tau_{ch} \equiv n_{ch} \Delta t\)
 with the corresponding integer \(n_{ch}\).
 \(F(t)\) is assumed to keep the same value for the interval
 \(\Delta t\), which is denoted as \(F_k = F(k \Delta t)\).
 The conditional probability \(p\) that \(F_{j+1}\) takes the same value
 as \(F_j\) is given by
\begin{equation}
p = e^{- \Delta t / \tau_f},
\label{eq:explicit-p}
\end{equation}
and the probability \(q\) that \(F_{k+1}\) is different from \(F_k\)
is therefore given by
\begin{equation}
q = 1 - p.
\label{eq:explicit-q}
\end{equation}
 The system is analyzed phenomenologically as follows:
\begin{itemize}
\item We introduce the variable \(s_k\) at a discretized time \(k \Delta t\)
 which takes two values \(\pm 1\).
\item \(s_k\) and \(F_k\) are initially set to
 \(s_0= + 1\) and \(F_0 = + H_0\), respectively.
\item \(s_k\) jumps from \(+1\) (\(-1\)) to \(-1\) (\(+1\))
 only if \(F_k\) continues to take the identical value
 \(-H_0\) (\(+H_0\)) for a time interval longer than \(n_{ch} \Delta t\).
\item \(s_k\) does not jump from \(+1\) (\(-1\)) to \(-1\) (\(+1\))
 even though \(F_k\) continues to take \(+ H_0\) (\(- H_0\))
 for any time interval longer than \(n_{ch} \Delta t\).
\end{itemize}
\subsection{MFPT \(\bar{\tau}\) through the channel}
 We first derive the exact expression for the MFPT
 \(\bar{\tau}\) through the channel with the phenomenological approach.
 In considering the time series having \(F_k\), \(\bar{\tau}\) is evaluated as
\begin{equation} 
\bar{\tau} = \left. \sum_{l} l \Delta t \sum_{0 \le k \le l}
 g_{k, l}^{(n_{ch})} q^k p^{l-k} \right|_{q=1-p}.
\label{eq:average-tau}
\end{equation}
 Here \(g_{k,l}^{(n_{ch})}\) is the number of the time sequences
 \(\{F_j\}\) for \(0 \le j < l\) satisfying that \(F_j\) changed its value
 \(k\) times in each \(\{F_j\}\) and \(s_j\) jumps from \(+1\) to \(-1\)
 for the first time at \(t=l \Delta t\).
\par
 Equation~(\ref{eq:average-tau}) is, furthermore, rewritten as
\begin{equation} 
\bar{\tau} = \left. \hat{T} Q_{n_{ch}}(q,p) \right|_{q=1-p}
\end{equation}
with the differential operator \(\hat{T}\) and the quantity \(Q_{n_{ch}}(q,p)\)
 defined by
\begin{equation}
\hat{T} \equiv \Delta t \left( q \frac{\partial}{\partial q} + 
 p \frac{\partial}{\partial p} \right)
\label{eq:T-operator}
\end{equation}
and
\begin{equation}
Q_{n_{ch}}(q,p) \equiv \sum_{l} \sum_{0 \le k \le l}
 g_{k, l}^{(n_{ch})} q^k p^{l-k}.
\label{eq:Q-q-p}
\end{equation}
 One should note that the \(q\)- and \(p\)-dependences in \(Q_{n_{ch}}\)
 are crucial and that \(q\) and \(p\) are considered to be independent
 in Eq.~(\ref{eq:Q-q-p}).
\par
 The explicit form of \(Q_{n_{ch}}(q,p)\)
 is then determined so as to satisfy the following conditions:
\begin{itemize}
\item In considering any length of time series giving \(F_k\),
 there exists a time interval of length \(n_{ch}\)
 in the last of the time series,
 where all the \(F_k\) take the same value \(-H_0\), i.e., the condition that
 \(s_k\) jumps from \(+1\) to \(-1\) is satisfied.
\item The condition for \(s_k\) to jump from \(+1\) to \(-1\)
 is not satisfied before the last time interval.
\end{itemize}
 One should note that the equality \(Q_n(1-p,p)=1\) holds for any \(n\),
 because the time interval described above always exists somewhere in a long
 time series.
 Particularly, for \(n=n_{ch}\), \(Q_{n_{ch}}(1-p,p)\) is obviously equal to
 the probability that \(s_j\) changes its sign, which must be unity
 for \(H_0 > H_c\).
\par
 As shown in Appendix~\ref{sec:form-Qn}, the explicit form of \(Q_n(q,p)\)
 is given by
\begin{equation}
Q_n(q,p) = \frac{(1-p)qp^{n-1}}{(1-p)^2 - q^2(1-p^{n-1})},
\label{eq:explicit-form-Qn}
\end{equation}
where the condition \(Q_n(1-p,p)=1\) is easily confined.
 Applying the operator (\ref{eq:T-operator}) to the explicit form
 (\ref{eq:explicit-form-Qn}) with \(n=n_{ch}\) yields the relation
\begin{eqnarray}
\bar{\tau} &=& \left. \hat{T} Q_{n_{ch}}(q, p) \right|_{q=1-p}
 = \Delta t \frac{ 2 - p^{n_{ch}-1}}{(1-p)p^{n_{ch}-1}} \nonumber \\
&=& \Delta t \frac{2-e^{\Delta t/\tau_f} e^{-\tau_{ch}/\tau_f}}
{ (1-e^{- \Delta t/\tau_f})e^{\Delta t/\tau_f} e^{- \tau_{ch}/\tau_f}},
\end{eqnarray}
where the last equality is obtained by using
 Eqs.~(\ref{eq:explicit-p}) and (\ref{eq:explicit-q}) with the relation
 \(\tau_{ch} = n_{ch} \Delta t\).
 The exact expression of \(\bar{\tau}\) is finally given by
\begin{equation}
\bar{\tau} = \tau_f \left( 2 e^{\tau_{ch}/\tau_f} - 1 \right)
\label{eq:solution-tau-vs-h0}
\end{equation}
in the limit of \(\Delta t \to 0\) by keeping \(\tau_{ch}\) constant.
 Equation (\ref{eq:solution-tau-vs-h0}) qualitatively agrees for
 \(\tau_{ch} / \tau_f \gg 1\)  with
 the result (\ref{eq:estimate-tau-vs-h0}).
\subsection{Distribution function \(P(\tau)\) for the passage time \(\tau\)}
 The distribution function \(P(\tau)\) for the passage time \(\tau\)
 through the channel \(s_{ch}\)
 is determined by solving the equation
\begin{equation}
P(\tau) = \left. \delta( \tau - \hat{T} ) Q_{n_{ch}}(q,p) \right|_{q=1-p},
\label{eq:distribution-func-tau}
\end{equation}
where \(\delta(x)\) is the delta function.
 The Laplace transform \({\cal L}[P](z)\) should be calculated in order to
 solve Eq.~(\ref{eq:distribution-func-tau}).
 By using the series expansion of \(Q_{n_{ch}}(q,p)\) given by
 Eq.~(\ref{eq:Q-q-p}), the Laplace transform of \(P(\tau)\) is obtained as
\begin{eqnarray}
{\cal L}[P](z) &\equiv& \int_0^{\infty} e^{-\tau z} P(\tau) d\tau =
 \left. e^{-z \hat{T}} Q_{n_{ch}}(q,p) \right|_{q=1-p}  \nonumber \\
&=& \left. \sum_{l} \sum_{0 \le k \le l} g_{k,l}^{(n_{ch})}
 ( e^{-z \Delta t} q)^k (e^{-z \Delta t} p)^{l-k} \right|_{q=1-p}.
\label{eq:formal-laplace-P}
\end{eqnarray}
 Equation (\ref{eq:formal-laplace-P}) implies that \({\cal L}[P](z)\) can be
 obtained by replacing \(q\) and \(p\) by  \(e^{-z \Delta t} q\) and
 \(e^{-z \Delta t} p\) in \(Q_{n_{ch}}(q,p)\), respectively, i.e.,
\begin{equation}
{\cal L}[P(\tau)](z) = \left. Q_{n_{ch}}( e^{-z \Delta t} q,
 e^{-z \Delta t} p ) \right|_{q=1-p}.
\label{eq:explicit-laplace-P}
\end{equation}
\par
 Substituting the explicit form (\ref{eq:explicit-form-Qn})
 for \(n=n_{ch}\) into Eq.~(\ref{eq:explicit-laplace-P}) yields
 the equation
\begin{equation}
{\cal L}[P(\tau)](z) =
\frac{
(\tau_f z + 1)e^{-(z + \tau_f^{-1})\tau_{ch}}
}{
\tau_f^2 z^2 + 2 \tau_f z + e^{-(z + \tau_f^{-1}) \tau_{ch}}
}
\label{eq:exact-laplace-P}
\end{equation}
in the limit of \(\Delta t \to 0\) by keeping \(\tau_{ch}\) constant.
 By applying the inverse Laplace transform to Eq.~(\ref{eq:exact-laplace-P}),
 the distribution function \(P(\tau)\) is analytically evaluated
 in the series expansion as
\begin{eqnarray}
P(\tau)&=&\tau_f^{-1}e^{-\tau/\tau_f}
\sum_{k=0}^\infty\theta(t_{k+1})\left.
\frac{(-x)^k}{k!}\frac{d^k}{dx^k}
\cosh\sqrt{x} \right|_{x=(t_{k+1})^2} \nonumber \\
 &=&\tau_f^{-1}e^{-\tau/\tau_f}
\Bigg[ \theta(t_1)\cosh(t_1)
-\theta(t_2)\frac{t_2\sinh(t_2)}{2}
+\theta(t_3)\frac{t_3{}^2\cosh(t_3) -t_3\sinh(t_3)}{8} \nonumber \\
&&
-\theta(t_4)\frac{ t_4{}^3\sinh(t_4)-3t_4{}^2\cosh(t_4)+3t_4\sinh(t_4)}{48}
+\cdots \Bigg],
\label{eq:distribution-P}
\end{eqnarray}
where \(t_k(\tau) \equiv (\tau - k \tau_{ch})/\tau_f\) and
 \(\theta(t)\) is the Heaviside function defined by
\begin{equation}
\theta(t) =
\left\{
\begin{array}{ccc}
1 & \mbox{for} & t \ge 0 \\
0 & \mbox{for} & t < 0
\end{array}
\right.
.
\label{eq:heaviside}
\end{equation}
 For details of the derivation of Eq.~(\ref{eq:distribution-P}),
 see Appendix \ref{sec:form-Pt}.
\par
\begin{figure}
\includegraphics[scale=0.5]{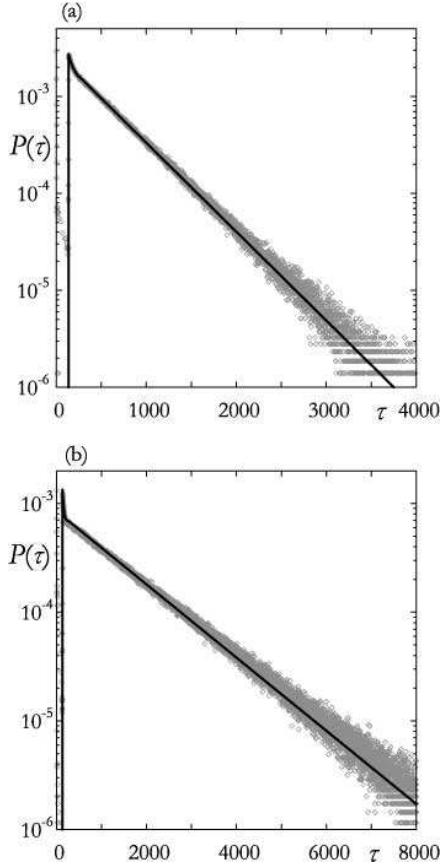}
\caption{\label{fig:distrib_Ptau} Distribution function \(P(\tau)\) of
 time passing through channels.
 The theoretical results (solid lines) are compared with the numerically
 evaluated distributions (symbols) for (a) \(\tau_f=130\)
 and (b) \(\tau_f=50\).
 The strength of DMN is set to be \(H_0=0.3852\) in both of the numerical
 simulations.
 \(\tau_{ch}\) corresponding to the applied DMN is
 estimated to be \(\tau_{ch} \simeq 134\)
 via the condition \(P(\tau) \simeq 0\) for \(\tau < \tau_{ch}\), where
 both (a) and (b) give the identical value of \(\tau_{ch}\).
 The expansion in Eq.~(\ref{eq:distribution-P}) is summed over for
 \(k \le 60\), i.e., \(k_c=60\).
}
\end{figure}

 Let us suppose to truncate the expansion (\ref{eq:distribution-P})
 at \(k=k_c\) for an arbitrary \(k_c\).
 It should be noted that Eq.~(\ref{eq:distribution-P}) gives the
 exact distribution for \(0 < \tau < k_c \tau_{ch}\)
 even though the truncation is executed,
 since all the terms individually
 include \(\theta(t_k)\) and so the terms for \(k>k_c\)
 do not contribute to \(P(\tau)\) for \(\tau < k_c \tau_{ch}\).
 The analytical result (\ref{eq:distribution-P}) is compared with
 the numerically evaluated distribution in Fig.~\ref{fig:distrib_Ptau}.
 One observes that the phenomenological analysis quantitatively explains
 the statistical property of passing through the channels.
 The characteristics obtained from the figure are summarized as follows.
\begin{itemize}
\item There exists a region where \(P(\tau)=0\) for \(\tau < \tau_{ch}\),
 which presents the minimal time of passing through the channels.
\item \(P(\tau)\) decreases exponentially for \(\tau \gg \tau_{ch}\),
 \(P(\tau) \propto e^{- \alpha \tau}\) with a constant \(\alpha\).
\item The rate \(\alpha\) increases as \(\tau_f\) is increased.
 This tendency is
 consistent with the fact that the probability of passing through channels
 increases as \(\tau_f\) is increased since DMN
 will often continue to take an identical value longer than \(\tau_{ch}\).
\end{itemize}
\par
 On the other hand, the expansion (\ref{eq:distribution-P})
 disagrees with the correct value
 in an exponential way for \(\tau > k_c \tau_{ch}\).
 Let us try to obtain the asymptotic solution of \(P(\tau)\)
 for \(\tau \gg \tau_{ch}\).
  Equation (\ref{eq:laplace-expand}) is approximated as
\begin{equation}
\langle e^{-z(\tau - \tau_{ch})} \rangle \simeq
 \frac{1}{1 + (\bar{\tau} - \tau_{ch}) z}
 \quad \mbox{for \(|z| \ll \tau_{ch}^{-1}\)},
\label{eq:approximate-laplace-P}
\end{equation}
where \(\bar{\tau}\) is MFPT given in Eq.~(\ref{eq:solution-tau-vs-h0}).
 The inverse Laplace transform of Eq.~(\ref{eq:approximate-laplace-P})
 is straightforwardly calculated to give
\begin{equation}
P(\tau) \simeq \frac{1}{\bar{\tau} - \tau_{ch}}
 \exp \left( - \frac{\tau - \tau_{ch}}{\bar{\tau} - \tau_{ch}} \right),
\end{equation}
which reveals that \(P(\tau)\) decreases exponentially with the damping rate
 \(\alpha = (\bar{\tau} - \tau_{ch})^{-1}\) for \(\tau \gg \tau_{ch}\).
\subsection{Fourier spectrum determined by the phenomenological analysis}
 We derive the Fourier spectrum of a time series \(s(t)\)
 by the phenomenological analysis to focus on the dynamical characteristics
 in the SRM phase.
 The Fourier spectrum \(I_x(\omega)\) is defined by
\begin{equation}
I_x(\omega) = \lim_{T \to \infty} \frac{1}{T} \left\langle \left| \int_0^T
 x(t) e^{-i \omega t} dt \right|^2 \right\rangle,
\label{eq:define-structure-function}
\end{equation}
i.e., the ensemble average of the Fourier transform
 of a time series \(x(t)\).
\par
 Let us first consider \(s_0(t) \equiv \mathrm{sgn}[s(t)]\).
 Then the time series \(s_0(t)\) is expressed as
\begin{equation}
s_0(t) = (-1)^{n-1}, \quad \mbox{for \(t_{n-1} \le t < t_{n}\)}
\label{eq:orbit-probability-model}
\end{equation}
with \(n \ge 1\), where \(t_n\) denotes the \(n\)th time to cross
 zero for \(s(t)\).
 Hereafter, \(t_0\) is set to be zero without loss of generality.
 By identifying that \(\tau_n \equiv t_{n}-t_{n-1}\) is independently
 distributed according to Eq.~(\ref{eq:distribution-P}),
 one obtains the Fourier spectrum of \(s_0(t)\) by the
 phenomenological analysis shown in Appendix \ref{sec:structure-func},
 in the form
\begin{equation}
I_{s_0} (\omega) = \frac{4}{\bar{\tau} \omega^2}
 \Re \left( \frac{ 1 - \langle e^{- i\omega \tau_n} \rangle}
{1+\langle e^{- i\omega \tau_n} \rangle} \right)
 = \frac{4}{\bar{\tau} \omega^2}
 \frac{ 1 - |\langle e^{- i\omega \tau_n} \rangle|^2 }
{|1+\langle e^{- i\omega \tau_n} \rangle |^2}
\label{eq:general-structure-function}
\end{equation}
where \(\Re(X)\) represents the real part of \(X\), and
\(\lim_{N\to\infty}\frac{t_N}{N}=\langle\tau_n\rangle=\bar{\tau}\) is used.
\par
\begin{figure}
\includegraphics[scale=0.5]{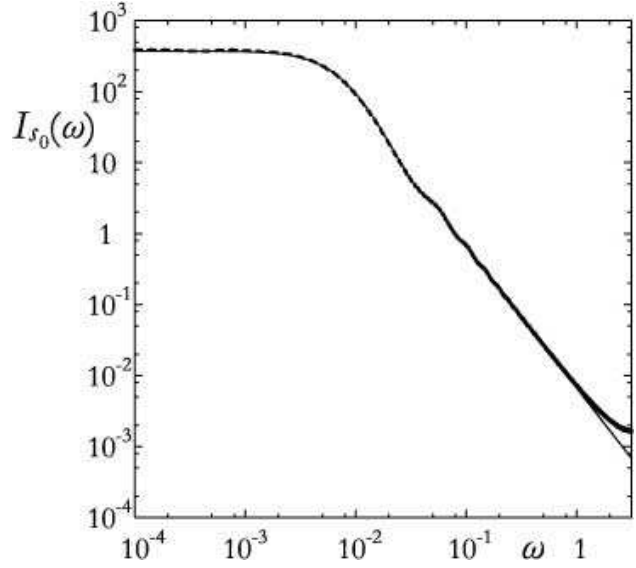}
\caption{\label{fig:power_normal} Fourier spectrum of \(s_0(t)\) obtained
 theoretically (solid line) and numerically (dashed line).
 \(\tau_f = 130\) is set for both numerical and theoretical results, and
 the strength of DMN is set to be \(H_0 = 0.3852\) in the numerical simulation.
 \(\tau_{ch}=134\) estimated in Fig.~\ref{fig:distrib_Ptau} is used
 for Eq.~(\ref{eq:get-structure-function}).
}
\end{figure}

 Substituting the explicit form of \(\langle e^{- i\omega \tau_n} \rangle\)
 given in Eq.~(\ref{eq:exact-laplace-P}) with \(z = i \omega\)
 into Eq.~(\ref{eq:general-structure-function}) yields
\begin{equation}
I_{s_0} ( \omega ) = \left( \frac{4 \tau_f}{ \bar{\tau} \omega} \right)
\frac{
\omega^3 \tau_f^3 +(4-e^{-2\tau_{ch}/\tau_f})\omega \tau_f
 - 2e^{-\tau_{ch}/\tau_f} (\omega \tau_f \cos \omega \tau_{ch}
 + 2 \sin \omega \tau_{ch} ) 
}{
(4+\omega^2 \tau_f^2) ( \omega^2 \tau_f^2 - 2 \omega \tau_f
 e^{-\tau_{ch}/\tau_f} \sin \omega \tau_{ch} + e^{-2\tau_{ch}/\tau_f})
}.
\label{eq:get-structure-function}
\end{equation}
 The above result is confirmed by comparing with
 the numerically evaluated Fourier spectrum for the normalized time series
 \(s_0(t)\) in Fig.~\ref{fig:power_normal}.
\par
 Let us finally modify the phenomenological analysis which
 is compatible with the numerically evaluated spectrum of the original time
 series \(s(t)\) without normalization.
 Instead of Eq.~(\ref{eq:orbit-probability-model}), let us define
\begin{equation}
\tilde{s}(t) = (-1)^{n-1}[1-a(t-t_{n-1})]
 \quad \mbox{for \(t_{n-1} \le t < t_{n}\)}
\end{equation}
with \(n \ge 1\), where \(a(\Delta t)\) incorporates the wave form of
 the time series passing through the channel and is assumed to be
 \(a(\Delta t)=0\) for \(\Delta t > \tau_{ch}\).
 Note that by setting  \(a(\Delta t)=0\) also for \(\Delta t \le \tau_{ch}\)
 the result of original phenomenological analysis is recovered.
 As shown in appendix \ref{sec:structure-func},
 the Fourier spectrum \(I_{\tilde{s}}(\omega)\) for \(\tilde{s}(t)\)
 as a modification to \(I_{s_0}(\omega)\) is obtained in the form
\begin{equation}
I_{\tilde{s}}(\omega) = I_{s_0}(\omega) \frac{1 + |\hat{a}(\omega)|^2
 + 2 \Re [\hat{a}(\omega)]}{4},
\label{eq:correct-structure-function}
\end{equation}
where
\begin{equation}
\hat{a}(\omega) \equiv 1 - i \omega \int_0^{\tau_{ch}} a(t) e^{- i\omega t} dt.
\end{equation}
 By approximating as \(a(\Delta t) = 1 + |s_{ch}|\) for
 \(0 < \Delta t < \tau_{ch}\),
 Eq.~(\ref{eq:correct-structure-function}) reduces to
\begin{equation}
I_{\tilde{s}}(\omega) = I_{s_0}(\omega) \left( \frac{1+ s_{ch}^2}{2}
 + \frac{1-s_{ch}^2}{2} \cos \omega \tau_{ch} \right).
\label{eq:explicit-correct-structure-function}
\end{equation}
\begin{figure}
\includegraphics[scale=0.5]{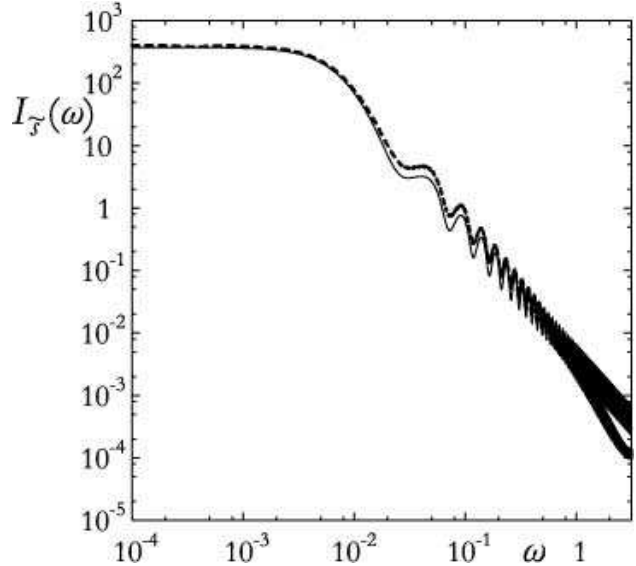}
\caption{\label{fig:power_full} Theoretically obtained Fourier spectrum
 (\ref{eq:explicit-correct-structure-function}) (solid line)
 with \(|s_{ch}| = 1/\sqrt{3}\)
 is compared with the numerical result (dashed line).
 \(\tau_f = 130\) is set for both numerical and theoretical results, and
 the strength of DMN is set to be \(H_0 = 0.3852\) in the numerical simulation.
 \(\tau_{ch}=134\) is used for Eq.~(\ref{eq:get-structure-function})
 which was estimated in Fig.~\ref{fig:distrib_Ptau}.
}
\end{figure}
 Equation (\ref{eq:explicit-correct-structure-function}) is confirmed
 by comparing with the numerically evaluated Fourier spectrum in
 Fig.~\ref{fig:power_full}.
\section{Concluding remarks}
\label{sec:concluding}
 In this paper, we used the model equation
 (\ref{eq:tdgl-external-field}) under the dichotomous Markov noise
 (DMN) \(F(t)\) with a finite correlation time
 in order to investigate the dynamics of the magnetization \(s(t)\)
 of the ferromagnet system driven by
 the magnetic field applied in one direction, where its strength is
 constant and only the direction temporally changes.
 It was found that the dynamics of \(s(t)\) show two kinds of
 motion, i.e., the symmetry-restoring motion (SRM) and
 the symmetry-breaking motion
 (SBM), which are respectively observed when \(H_0\) is above and below
 the critical value \(H_c\).
 The transition line between SRM and SBM was determined
 only by the strength of the applied DMN and is independent of
 the correlation time \(\tau_f\) of DMN.
 By observing the distribution functions of \(s(t)\) for SRM and SBM,
 the ensemble average of
 \(s(t)\) discontinuously changes at \(H_0 = H_c\).
 These results are quite different from those in the system driven by
 a periodically oscillating field \cite{fujisaka1}.
\par
 We then discussed the mean first passage time (MFPT)
 slightly above \(H_c\)
 and found that it depends on \(H_0-H_c\) and \(\tau_f\) as
\begin{equation}
\bar{\tau}\simeq \tau_f \exp \left[ \frac{C}{\tau_f (H_0 - H_c)^{1/2}} \right].
\end{equation}
 This anomalous characteristics has a form similar to that of the
 average duration between neighboring phase slips of the phase difference
 in the phase synchronization observed in coupled chaotic
 systems~\cite{rosenblum}.
 Furthermore, a phenomenological approach was proposed to analytically discuss
 the statistical characteristics for \(H_0 > H_c\).
 By obtaining the probability \(Q_n(q,p)\) of \(s(t)\) passing through
 the channel, the MFPT \(\bar{\tau}\),
 the distribution function \(P(\tau)\) of the passage time \(\tau\)
 and the Fourier spectrum \(I(\omega)\) of the time series were
 obtained by using the phenomenological analysis.
 The statistics obtained by the phenomenological analysis
 were found to be not only in qualitative but also in quantitative
 agreement with the numerically obtained results.
\par
 In closing the paper, it is worth noting that the effect of
 DMN on nonlinear dynamical systems is generally quite different from
 that of Gaussian noise and DMN
 produces a new dynamical response of the systems.
 It is highly desired to examine the statistical characteristics
 obtained in this paper in laboratory experiments, and also
 in other numerical simulations, e.g., Monte Carlo simulations.
\begin{acknowledgments}
 The authors thank N.~Tsukamoto, N.~Fujiwara, and H.~Hata
 for valuable comments.
 This study was partially supported by Grant-in-Aid
 for Scientific Research (C)
 of the Ministry of Education, Culture, Sports, Science, and Technology,
 and the 21st Century COE Program ``Center of Excellence for
 Research and Education on Complex Functional Mechanical Systems''
 at Kyoto University.
\end{acknowledgments}
\appendix
\section{Explicit form of \(Q_n(q,p)\)}
\label{sec:form-Qn}
First of all, notice that each sample path of $F_k$ ($k=0, 1, 2,
\cdots$) corresponds to a symbol sequence of $\{+,-\}$.  Let $w$ be a
symbol sequence of $\{+, -\}$, which is referred to as a string.
 For given sets $A$ and $B$ of strings, let $AB$ be the set of all strings
expressed as $ab$ for $a\in A$ and $b\in B$, where $ab$ denotes the
concatenated string of $a$ followed by $b$.  In the following, the set
composed of only one string $w$ will be simply expressed as $w$.
Furthermore, let $(A)_n$ be the set of all strings expressed as
\begin{equation}
  a_1a_2\cdots a_k
\end{equation}
with $a_i\in A$ and $0\le k\le n$, where $k=0$ means the zero length
string, including the case $a_i=a_j$ for $i\ne j$.  Then, with this
notation, every possible sequence starting with $+$ and terminating
with successive $-$'s of length $n$ $(n\ge2)$, which firstly appears in
that string, can be expressed as
\begin{equation}
  S_n\equiv (+(+)_\infty-(-)_{n-2})_\infty+(+)_\infty
  \overbrace{--\cdots-}^{n \mbox{ symbols}}.
\end{equation}

For a given set $S$ of strings, let us define its ``probability''
as a function of $p$ and $q$ by
\begin{equation}
  P(q, p; S) \equiv \sum_{w\in S} P_w(p, q)
\end{equation}
where $P_w(p, q) \equiv p^{k(w)}q^{l(w)}$ denotes the probability
 for a string $w$, and
$k(w)$ and $l(w)$ denote the numbers of pairs of identical symbols
($--$, $++$) and different symbols ($-+$, $+-$) appearing in $w$,
respectively.
For given strings $w_1$ and $w_2$, obviously, the identity
\begin{equation}
  P_{w_1w_2}(q, p) =  P_{w_1\sigma}(q, p) P_{w_2}(q, p)
\end{equation}
holds with $\sigma$ being the first symbol in $w_2$, and we
call $w_1\sigma$ and $w_2$ a decomposition of the string $w\equiv
w_1w_2$.  By noting that
\begin{equation}
  Q_n(q, p) = P(q, p; S_n)
\end{equation}
and considering decompositions of each element in $S_n$,
$Q_n(q, p)$ can be expressed as
\begin{equation}
  \label{eq:q_n}
  Q_n(q, p) = P(q, p; S'_n)P(q, p; R)p^{n-1},
\end{equation}
where 
\begin{equation}
  S'_n\equiv(+(+)_\infty-(-)_{n-2})_\infty+
\end{equation}
and
\begin{equation}
    R\equiv +(+)_\infty -.
\end{equation}
Since each element of $S'_n$ can be uniquely decomposed into a
multiple of elements in
\begin{equation}
  S''_n\equiv +(+)_\infty-(-)_{n-2}+
\end{equation}
and inversely every multiple of elements in $S''_n$ uniquely
corresponds to an element in $S'_n$ as its decomposition, we obtain
\begin{eqnarray}
  P(q, p; S'_n) &=& \sum_{j=0}^\infty \left[P(q, p; S''_n)\right]^j\\
  \label{eq:p-s_n}
  &=& \frac{1}{1-P(q, p; S''_n)}.
\end{eqnarray}
Finally, $P(q, p; R)$ and $P(q, p; S''_n)$ are calculated as
\begin{equation}
  P(q, p; R) = q\sum_{j=0}^\infty p^j=\frac{q}{1-p}
\end{equation}
and
\begin{equation}
  P(q, p; S''_n)=\sum_{j=0}^\infty p^jq\sum_{i=0}^{n-2}p^iq=
  q^2\frac{1-p^{n-1}}{(1-p)^2}, 
\end{equation}
which yield Eq.~(\ref{eq:explicit-form-Qn}) together with 
(\ref{eq:q_n}) and (\ref{eq:p-s_n}).
\section{Derivation of the probability distribution function \(P(\tau)\)}
\label{sec:form-Pt}
 Let us evaluate the inverse Laplace transform of
Eq.~(\ref{eq:exact-laplace-P}) in order to derive the explicit form
of \(P(\tau)\).
 For simplicity, the time is rescaled as \(\tau_f = 1\).
 Then Eq.~(\ref{eq:exact-laplace-P}) is rewritten as
\begin{equation}
  \langle e^{-z \tau} \rangle =
  \frac{(1+z) e^{-(1+z) \tau_{ch}}}{(1+z)^2 - (1-e^{-(1+z)\tau_{ch}})},
\end{equation}
which reads
\begin{equation}
  {\cal L}[P(\tau+\tau_{ch})]=
  \langle e^{-z(\tau - \tau_{ch})} \rangle =
  \frac{e^{-\tau_{ch}}}{1+z - \frac{1-e^{-(1+z)\tau_{ch}}}{1+z}} =
  e^{-\tau_{ch}}
  \sum_{n=0}^{\infty} \frac{\left(1-e^{-(1+z)\tau_{ch}} \right)^n}
  {(1+z)^{2 n+1}}.
\label{eq:laplace-expand}
\end{equation}
\par
Repeatedly applying the formula
\begin{equation}
  {\cal L}^{-1} \left[ e^{-(1+z)\tau_{ch}} \hat{f}(z) \right] =
  e^{-\tau_{ch}} f(\tau-\tau_{ch}) =
  e^{-\tau_{ch}} e^{- \tau_{ch}\frac{d}{d\tau}} f(\tau)
  =e^{-\tau_{ch}} e^{- \tau_{ch}\frac{d}{d\tau}} {\cal L}^{-1}[\hat{f}(z)]
\end{equation}
with \(f(\tau)\equiv{\cal L}^{-1}[\hat{f}(z)]\), one obtains
\begin{eqnarray}
  {\cal L}^{-1}\left[\frac{\left(1-e^{-(1+z)\tau_{ch}}\right)^n}
    {(1+z)^{2 n+1}}\right]
  &=&(1-e^{-\tau_{ch}}e^{-\tau_{ch}\frac{d}{d\tau}})^n
  {\cal L}^{-1}\left[\frac{1}{(1+z)^{2n+1}}\right] \nonumber \\
\nonumber  \\
  &=&(1-e^{-\tau_{ch}}e^{-\tau_{ch}\frac{d}{d\tau}})^n
  \frac{\tau^{2n}}{(2n)!}e^{-\tau}\theta(\tau),
\end{eqnarray}
where $\theta(\tau)$ denotes the Heaviside function 
Eq.~(\ref{eq:heaviside}), and the formula
\begin{equation}
  {\cal L}^{-1} \left[ \frac{1}{(1+z)^{m}} \right] =
  \frac{\tau^{m-1}}{(m-1)!} e^{-\tau} \theta(\tau)
\end{equation}
for any positive integer \(m\) was used.
Thus, the inverse Laplace transform of
 Eq.~(\ref{eq:laplace-expand}) reads
\begin{eqnarray}
  P(\tau+\tau_{ch})
  &=&
  e^{-\tau_{ch}}\sum_{n=0}^{\infty}
  \left(1-e^{-\tau_{ch}}e^{-\tau_{ch}\frac{d}{d\tau}} \right)^n
  \frac{\tau^{2n}}{(2n)!} e^{-\tau}\theta(\tau) \nonumber \\
  &=&e^{-(\tau+\tau_{ch})} \sum_{n=0}^{\infty}
  \left(1 - e^{- \tau_{ch}\frac{d}{d\tau}} \right)^n
  \frac{\tau^{2n}}{(2n)!} \theta(\tau),
  \label{eq:p-tau-1}
\end{eqnarray}
where the identity
\begin{equation}
  e^{-\tau_{ch}} e^{-\tau_{ch} \frac{d}{d\tau}}
  e^{-\tau} = e^{-\tau} e^{-\tau_{ch} \frac{d}{d\tau}}
\end{equation}
was applied.
By noting the identity
\begin{equation}
  \left(1 - e^{-\tau_{ch}\frac{d}{d\tau}} \right)^n
  =\sum_{k=0}^n\frac{(-e^{-\tau_{ch}\frac{d}{d\tau}})^k}{k!}
  \left.\frac{d^kx^n}{dx^k}\right|_{x=1}
  =\left.\sum_{k=0}^\infty\frac{e^{-k\tau_{ch}\frac{d}{d\tau}}(-1)^k}{k!}
  \left(\frac{d}{dx}\right)^kx^n\right|_{x=1}
\end{equation}
Eq.~(\ref{eq:p-tau-1}) is further simplified as
\begin{eqnarray}
  P(\tau+\tau_{ch})&=&\left.e^{-(\tau+\tau_{ch})}
    \sum_{n=0}^\infty\sum_{k=0}^\infty e^{-k\tau_{ch}\frac{d}{d\tau}}
    \frac{(-1)^k}{k!}\left(\frac{d}{dx}\right)^kx^n
    \frac{\tau^{2n}}{(2n)!}\theta(\tau)
  \right|_{x=1} \nonumber \\
  &=&\left.e^{-(\tau+\tau_{ch})}\sum_{k=0}^\infty
    e^{-k\tau_{ch}\frac{d}{d\tau}}
    \frac{(-1)^k}{k!}\left(\frac{d}{dx}\right)^k
    \cosh(\sqrt{x}\tau)\theta(\tau)\right|_{x=1} \nonumber \\
  &=&\left.e^{-(\tau+\tau_{ch})}\sum_{k=0}^\infty
    e^{-k\tau_{ch}\frac{d}{d\tau}}\theta(\tau)
    \frac{(-\tau^2)^k}{k!}\left(\frac{d}{dx}\right)^k
    \cosh\sqrt{x}\right|_{x=\tau^2} \nonumber \\
  &=&\left.e^{-(\tau+\tau_{ch})}\sum_{k=0}^\infty\theta(t_k)
    \frac{(-x)^k}{k!}\frac{d^k}{dx^k}\cosh\sqrt{x}\right|_{x=t_k^2},
\label{eq:p-tau-fin}
\end{eqnarray}
where \(t_k(\tau)\equiv\tau-k\tau_{ch}\) and the formula \(\sum_{n=0}^{\infty}
\frac{x^n}{(2n)!} = \cosh \sqrt{x}\) was used.  After the replacement
\(\tau \to \tau - \tau_{ch}\) and the rescaling of time as \(\tau \to
\tau/\tau_f\) and  \(\tau_{ch} \to \tau_{ch}/\tau_f\),
 Eq.~(\ref{eq:p-tau-fin}) reduces to Eq.~(\ref{eq:distribution-P}).
\section{Derivation of the Fourier Spectrum \(I(\omega)\)}
\label{sec:structure-func}
In this appendix, we derive the Fourier spectrum of the time series
 of the magnetization \(s(t)\) for the phenomenological analysis.
 Let \(\tau_1, \tau_2, \tau_3, \cdots\) be a sequence of mutually
independent random variables having the probability density
\(P(\tau)\) obeying the condition \(P(\tau)=0\) for
\(\tau<\tau_{ch}\), i.e., \(\tau_k\ge\tau_{ch}\).
 We introduce the variable
\begin{equation}
  \tilde{s}(t)=(-1)^n[1-a(t-t_{n-1})] \mbox{  for  } t_{n-1}\le t <t_n,
\end{equation}
where \(t_n\equiv\sum_{k=1}^N\tau_k\) and \(a(\Delta t)\) is a
function satisfying \(a(\Delta t)=0\) for \(\Delta t>\tau_{ch}\).  We
assume that the time series of the magnetization \(s(t)\) is approximately
expressed by \(\tilde{s}(t)\) with an appropriate form of \(a(t)\).  The
Fourier transform of \(\tilde{s}(t)\) follows
\begin{eqnarray}
  \int_0^{t_N} \tilde{s}(t) e^{-i \omega t} dt &=& \sum_{n=1}^{N} (-1)^{n-1} 
  \int_{t_{n-1}}^{t_n}[1-a(t-t_{n-1})]e^{-i\omega t}dt \nonumber \\
  &=&\sum_{n=1}^N(-1)^{n-1}e^{-i\omega t_{n-1}}\left[
    \int_0^{\tau_n}e^{-i\omega t}dt-\int_0^{\tau_{ch}}a(t)e^{-i\omega t}dt
  \right] \nonumber \\
  &=&\sum_{n=1}^N(-1)^{n}e^{-i\omega t_{n-1}}\frac{e^{-i\omega \tau_n}
    -\hat{a}(\omega)}{i\omega},
\label{eq:fourier-probability-model}
\end{eqnarray}
where
\begin{equation}
   \hat{a}(\omega)\equiv 1-i\omega\int_0^{\tau_{ch}}a(t)e^{-i\omega t}dt.
\end{equation}
Considering the absolute square of
Eq.~(\ref{eq:fourier-probability-model}) and then taking the ensemble
average, we obtain
\begin{eqnarray}
  \lefteqn{\omega^2\left| \int_0^{t_N} \tilde{s}(t) 
      e^{-i \omega t} dt \right|^2 = 
    \sum_{n=1}^N\left|e^{-i\omega \tau_n}-\hat{a}(\omega)\right|^2}
  \nonumber\\
  && +2\Re\left[\sum_{1 \le m < n \le N}(-1)^{n-m}
  e^{-i \omega \sum_{k=m+1}^{n-1} \tau_k}
  \left(e^{-i\omega \tau_n}-\hat{a}(\omega)\right)
  \left(1 - e^{-i\omega \tau_m}\hat{a}^*(\omega)\right)\right]
\label{eq:absolute-fourier-probability}
\end{eqnarray}
and
\begin{eqnarray}
  \lefteqn{\frac{\omega^2}{N}\left\langle\left|
        \int_0^{t_N} \tilde{s}(t) e^{-i \omega t} dt \right|^2
    \right\rangle=
    1+|\hat{a}(\omega)|^2 - 2\Re\left[
      \langle e^{-i\omega\tau_n}\rangle \hat{a}^*(\omega)\right]}
    \nonumber\\
    && + 2N^{-1}\Re\left[\sum_{1 \le m < n \le N}(-1)^{n-m}
    \langle e^{-i\omega\tau_k}\rangle^{n-m-1}
    \left(\langle e^{-i\omega\tau_k}\rangle-\hat{a}(\omega)\right)
    \left(1-\langle e^{-i\omega\tau_k}\rangle \hat{a}^*(\omega)\right)
  \right].
  \label{eq:fourier-1}
\end{eqnarray}
 Since  \(|\langle e^{-i\omega\tau_n}\rangle|<1\),  in
the limit of \(N\to\infty\), the last term in
Eq.~(\ref{eq:fourier-1}) reads
\begin{eqnarray}
  \lefteqn{-2\Re\left[ \sum_{k=0}^\infty
      (-1)^k\langle e^{-i\omega\tau_n}\rangle^k
      \left(\langle e^{-i\omega\tau_n}\rangle -\hat{a}(\omega)\right)
      \left(1-\langle e^{-i\omega\tau_n}\rangle \hat{a}^*(\omega)\right)
  \right]}
  \nonumber\\
  &&=-2\Re\left[\frac{\langle e^{-i\omega\tau_n}\rangle
      \left(1+|\hat{a}(\omega)|^2\right)-
      \hat{a}(\omega)-\langle e^{-i\omega\tau_n}\rangle^{2}
      \hat{a}^*(\omega)}{1+\langle e^{-i\omega\tau_n}\rangle}\right],
\end{eqnarray}
which leads to
\begin{equation}
  \lim_{N\to\infty}
  \frac{1}{N}\left\langle\left|
      \int_0^{t_N} \tilde{s}(t) e^{-i \omega t} dt \right|^2
  \right\rangle=\omega^{-2}
  \Re\left[\frac{1-\langle e^{-i\omega\tau_n}\rangle}{1+
      \langle e^{-i\omega\tau_n}\rangle}\right]\left(
      1+|\hat{a}(\omega)|^2 + 2\Re[\hat{a}(\omega)]\right).
\end{equation}
 Noting that
\(\lim_{N\to\infty}\frac{t_N}{N}=\langle\tau_n\rangle=\bar{\tau}\), we obtain
\begin{equation}
  I_{\tilde{s}}(\omega)=I_{s_0}(\omega)
  \frac{1+|\hat{a}(\omega)|^2+2\Re[\hat{a}(\omega)]}{4},
\end{equation}
where 
\begin{equation}
  I_{s_0}(\omega)\equiv
  \frac{4}{\omega^2\bar{\tau}}
  \Re\left[\frac{1-\langle e^{-i\omega\tau_n}\rangle}{1+
      \langle e^{-i\omega\tau_n}\rangle}\right]
  \label{eq:power-spectrum-1}
\end{equation}
corresponds to the case that 
\(a(\Delta t)=0\) for all \(\Delta t\ge0\) and thus \(s_0(t)=(-1)^n\)
for \(t_{n-1}\le t<t_n\).

\end{document}